\documentclass[epsfig,useAMS,usegraphicx,usenatbib]{mn2e}

\title[HESS J1616--508: likely powered by PSR J1617--5055]
{HESS J1616--508: likely powered by PSR J1617--5055}
\author[R. Landi et al.]
{R.~Landi$^{1}$\thanks{E-mail:landi@iasfbo.inaf.it},
A.~De Rosa$^2$,
A.J.~Dean$^3$, L.~Bassani$^1$, P.~Ubertini$^2$, A.J.~Bird$^3$\\
$^1$ INAF -- IASF Bologna, Via P. Gobetti 101, I--40129 Bologna, Italy\\
$^2$ INAF -- IASF Rome, Via Fosso del Cavaliere 100, I--00133 Roma, Italy\\
$^3$School of Physics and Astronomy, University of Southampton,
        SO17 1BJ, Southampton, U.K.}

\begin{document}

\date{}

\pagerange{\pageref{firstpage}--\pageref{lastpage}} \pubyear{2007}

\maketitle

\label{firstpage}

\begin{abstract}
HESS J1616--508 is one of the brightest emitters in the TeV sky. Recent observations with the
IBIS/ISGRI telescope on board the \emph{INTEGRAL} spacecraft have revealed that a young, nearby
and energetic pulsar, PSR J1617--5055, is a powerful emitter of soft $\gamma$-rays in the 
20--100 keV domain. In this paper we present an analysis of all available data from the 
\emph{INTEGRAL}, \emph{Swift}, \emph{Beppo}SAX and \emph{XMM-Newton} telescopes with a view 
to assessing the most likely counterpart to the HESS source. We find that the energy source that fuels 
the X/$\gamma$-ray emissions is derived from the pulsar, both on the basis of the 
positional morphology, the timing evidence and the energetics of the system. Likewise the 1.2$\%$ of the 
pulsar's spin 
down energy loss needed to power the 0.1--10 TeV emission is also fully consistent with other
HESS sources known to be associated with pulsars. The relative sizes of the X/$\gamma$-ray
and VHE sources are consistent with the expected lifetimes against synchrotron and Compton losses
for a single source of parent electrons emitted from the pulsar. We find that no other known object
in the vicinity could be reasonably considered as a plausible counterpart to the 
HESS source. We conclude that there is good evidence to assume that the HESS J1616--508 
source is driven by PSR J1617--5055 in which a combination of synchrotron and inverse Compton
processes combine to create the observed morphology of a broad-band emitter from keV to TeV 
energies.
\end{abstract}


\section{Introduction}
The High Energy Stereoscopic System (HESS), a ground-based Cherenkov array telescope,
has recently reported results of a first sensitive survey of the Galactic plane with an
average flux  
sensitivity of a few percent of the Crab at energies above 100 GeV, revealing the 
existence of a new population of TeV sources 
(Aharonian et al. 2005a, 2006a).
The detection of a significant number of such sources
has important astrophysical implications for our understanding of 
cosmic particle acceleration. 
Different types of Galactic sources are known to be cosmic particle accelerators and potential 
sources of high energy $\gamma$-rays: isolated pulsars/pulsar wind nebulae (PWNs), supernova 
remnants (SNRs), star forming regions, binary systems with a collapsed object like a microquasar 
or a pulsar etc. 
Several of these new TeV emitters have no obvious counterpart at other energies
thus making the search for X and soft $\gamma$-ray emission of key importance 
to disentangle the mechanisms involved in the different emitting regions and, in turn, to
understand the nature of these objects. 

HESS J1616--508, with a flux of $(4.3\pm0.2)\times10^{-11}$ erg cm$^{-2}$ s$^{-1}$
above 200 GeV,
is one of the brightest of these new HESS sources (Aharonian et al. 2005a; 
Aharonian et al. 2006a). Its TeV spectrum is well modelled with a power law 
having a photon index $\Gamma = 2.35\pm 0.06$.
The source is spatially extended, with a rather circular shape, 
an angular diameter of $\sim$16$^{\prime}$ and it does not seem to have any counterpart at 
other wavelengths. Previous observations of this region performed in X-rays by 
\emph{Chandra} (Vink 2004) did not detect any extended X-ray source associated with the TeV 
emission.
HESS J1616--508 is located in a complex region containing two known SNRs, RCW 
103 (G332.4--0.4) and Kes 32 (G332.4+0.1), which  
are not spatially coincident with the HESS extension, being located 
13$^{\prime}$ and 17$^{\prime}$ away, respectively.
The first is a young shell type SNR, 
quite bright in X-rays (0.5--10 keV flux $\sim$$\times10^{-12}$ erg cm$^{-2}$ s$^{-1}$)
and with an absorbed soft spectrum (Becker \& Aschenbach 2002); it is 
characterized as having an X-ray point source, 1E 161348--5055.1, at its centre. Since its 
discovery,
this point source has been shown to be a peculiar object: recent \emph{XMM-Newton} data 
indicates the presence 
of a long period and extreme variability indicating that it may be either an X-ray binary 
in an eccentric orbit or a peculiar type of magnetar (De Luca et al. 2006).
The second SNR, Kes 32, is only detected at harder X-ray energies
(with a 1.5--5 keV flux of $(1-2)\times10^{-12}$ erg 
cm$^{-2}$ s$^{-1}$) because of a large interstellar absorption 
($\sim$$4\times10^{22}$ cm$^{-2}$) and is
characterized by a thermal spectrum of $kT$ $\sim$1 keV (Vink 2004).

In the vicinity (9$^{\prime}$ away) of HESS J1616--508 a young X-ray emitting pulsar
PSR J1617--5055 is also present; it was serendipitously discovered during 
\emph{ASCA} observation of RCW 103 (Torii et al. 1998) and it is, assuming a magnetospheric outer gap 
model, predicted to emit in the TeV regime (Hirotani 2001).
Two other pulsars (PSR J1616--5109 and PSR J1614--5048) are further away and less likely
to produce high energy $\gamma$-rays. 
Recently, HESS J1616--508 has been observed with the \emph{Suzaku} X-ray Imaging Spectrometer (XIS).
No positive detection of X-ray emission has been found 
down to a 2--10 keV limiting flux of $3.1\times10^{-13}$ erg cm$^{-2}$ s$^{-1}$ (Matsumoto et al. 
2007).
Similarly, \emph{XMM-Newton} data, analysed by the same authors, set an upper limit 
to the X-ray flux 
three times higher than the \emph{Suzaku} one. 
It follows that the source, bright in the TeV $\gamma$-rays, is faint in the X-ray band,
raising the possibility that we are dealing with a ``dark particle accelerator''.

In this work we present the analysis of observations of HESS J1616--508 performed 
with the \emph{INTEGRAL} (Winkler et al. 2003) and \emph{Swift} 
(Gehrels et al. 2004) satellites. In addition, we also use 
\emph{Beppo}SAX and 
\emph{XMM-Newton} archival data that cover the TeV extension/region. 
All these data are combined and discussed in order to 
assess the most likely counterpart to the HESS source and to understand its nature.

\begin{figure}
\includegraphics[width=1.0\linewidth]{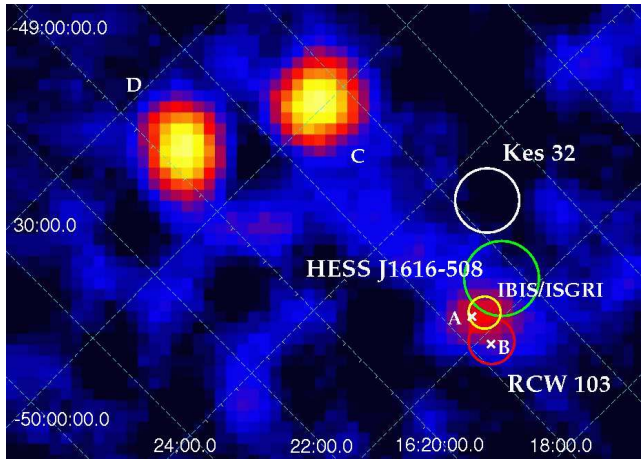}
\caption{\emph{INTEGRAL} 18--60 keV energy band image of the region surrounding HESS J1616--508.
The green circle represents the TeV extension while PSR J1617--5055, within the
IBIS/ISGRI error box (smaller yellow circle), is labelled as A. Also plotted is the position of 1E
161348--5055.1 (B), the source located at the centre of the SNR RCW 103 (red circle) and the SNR 
Kes 32 (white circle). Two nearby sources, AX J161929--4945 (C) and IGR J16207--5129 (D) are also
shown.}
\label{fig:integral}
\end{figure}

\section{\emph{INTEGRAL} observation}

The region surrounding HESS J1616--508 has been covered by the third \emph{INTEGRAL} 
IBIS/ISGRI survey (Bird et al. 2007),
which consists of several pointings from the beginning of the mission (November 2002)
up to April 2006. The total exposure on this region is of $\sim$2 Ms. ISGRI images for
each available pointing were generated in various energy bands using
the ISDC offline scientific analysis software OSA (Goldwurm et al. 2003) version 5.1.
The individual images were then combined to produce a mosaic of the entire sky to enhance the
detection significance using the system described in details by Bird
et al. (2004, 2007). Figure~\ref{fig:integral} shows the 18--60 keV energy band image
of the region surrounding HESS J1616--508: a clear excess is
detected with a significance of $\sim$7.4$\sigma$ at a position 
corresponding to 
R.A.=16$^{\rm h}$17$^{\rm m}$ 12$^{\rm s}$.72 and Dec=--50$^{\circ}$ 56$^{\prime}$ 31$^{\prime \prime}$.20
(J2000.0) with a positional uncertainty of 3$^{\prime}.5$.
Although the IBIS/ISGRI excess is compatible with the HESS source, within the
ISGRI error box (smaller yellow circle in the figure) there is only the pulsar PSR J1617--5055.
Fluxes for spectral analysis were extracted from narrow-band mosaics of
all pointings in the third \emph{INTEGRAL} IBIS/ISGRI survey which are relevant to the sky region
containing HESS J1616--508.
Here and in the following, spectral analysis have been performed using XSPEC v. 12.2.1;
quoted errors correspond to 90$\%$ confidence level for one interesting parameter 
($\Delta\chi^{2}=2.71$).
A simple power law model in the 17--300 keV energy band
provides a good fit to the IBIS/ISGRI data ($\chi^{2}/\nu=8.1/12$)
and a photon index 
$\Gamma=1.91_{-0.35}^{+0.40}$ combined with a
20--100 keV flux of $1.37\times$ 10$^{-11}$ erg cm$^{-2}$ s$^{-1}$.
As there is no \emph{INTEGRAL} detection of the TeV object,
we can derive from the IBIS/ISGRI mosaics a 2$\sigma$ upper limit
to the flux of any undetected counterpart: $1.5\times10^{-12}$ and 
$3.8\times10^{-12}$ erg cm$^{-2}$ s$^{-1}$ in the 20--40 and 
40--100 keV energy band, respectively. To conclude, \emph{INTEGRAL}
provides evidence for soft $\gamma$-ray emission from the nearest pulsar
to HESS J1616--508, which was already suggested 
to be a likely $\gamma$-ray emitter; alternatively, stringent upper limits
on the emission of the TeV object are provided in the 20--100 keV energy band.

\section{\emph{Swift}/XRT observation and data analysis}

HESS J1616--508 was observed with XRT (X-ray Telescope, 0.2--10 keV) on board the 
\emph{Swift} satellite (Gehrels et al. 2004) on several occasions during the period
June 2006 -- March 2007; the exposures span from $\sim$4 to $\sim$6 ks. 
Data reduction was performed using the XRTDAS v. 2.0.1 standard data pipeline package
({\sc xrtpipeline} v. 0.10.6), in order to
produce screened event files. All data were extracted only in the Photon Counting
(PC) mode (Hill et al. 2004), 
adopting the standard grade filtering (0--12 for PC) according to the XRT nomenclature.
All available XRT images of HESS J1616--508 in the 0.3--10 keV band were analysed with the 
XIMAGE package (v. 4.3). 
Within the HESS extension, we find only one source, which is detected in all observations,
at a confidence level of 3--4$\sigma$, and is located at
R.A.=16$^{\rm h}$15$^{\rm m}$ 50$^{\rm s}$.94 and
Dec=--50$^{\circ}$ 49$^{\prime}$ 26$^{\prime \prime}$.39 (J2000.0), with a 90\% uncertainty of 
5$^{\prime \prime}$; it was not seen by \emph{Suzaku} because it was outside its field of view. 
This object (with a 2--10 keV flux of $3.8\times10^{-13}$ erg cm$^{-2}$ 
s$^{-1}$ estimated assuming a Crab-like spectrum) has a  soft spectrum, being detected at 
the lowest XRT energies and 
also being reported as a ROSAT/HRI source; it has a USNO-B1.0 (Monet et al. 1999) and 2MASS (Two 
Micron All Sky Survey, Skrutskie et al. 2006) counterpart at R.A.=16$^{\rm h}$15$^{\rm m}$ 
50$^{\rm s}$.99 and
Dec=--50$^{\circ}$ 49$^{\prime}$ 24$^{\prime \prime}$.9 (J2000.0), with magnitudes $K$ 
$\simeq$9.0 and 
$R$ $\simeq$12.1; the $R-K$ colour index of $\sim$3 implies extinction in the source direction.

The random chance probability of finding an object as bright as this \emph{Swift} source inside the 
HESS extension is 0.2, making the association with the TeV emitter doubtful. 
This finding suggests that other possible counterparts may also be searched outside 
the TeV extension.

\section{\emph{Beppo}SAX/MECS observations and data analysis}
\begin{figure}
\includegraphics[width=1.0\linewidth]{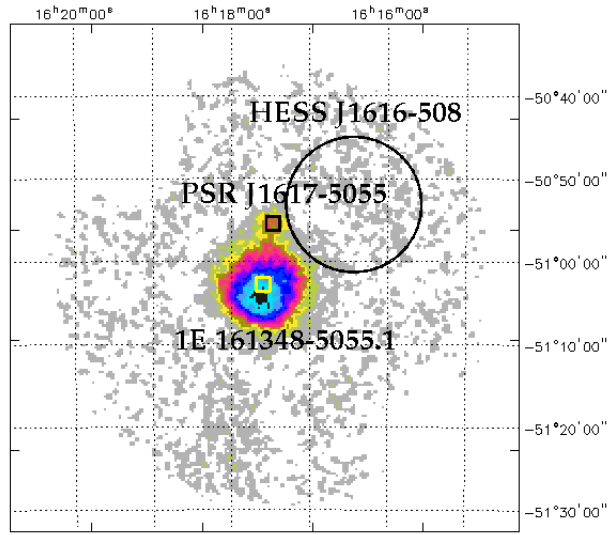}
\caption{MECS 2--4 keV image of the region surrounding 1E 161348--5055.1. The black
circle describes the HESS J1616--508 extension, while the positions of 1E 161348--5055.1
and PSR J1617--5055 are given by boxes. The SNR RCW 103 around 1E 161348--5055.1 is clearly visible.}
\label{fig:mecs1}
\end{figure}

\begin{figure}
\includegraphics[width=1.0\linewidth]{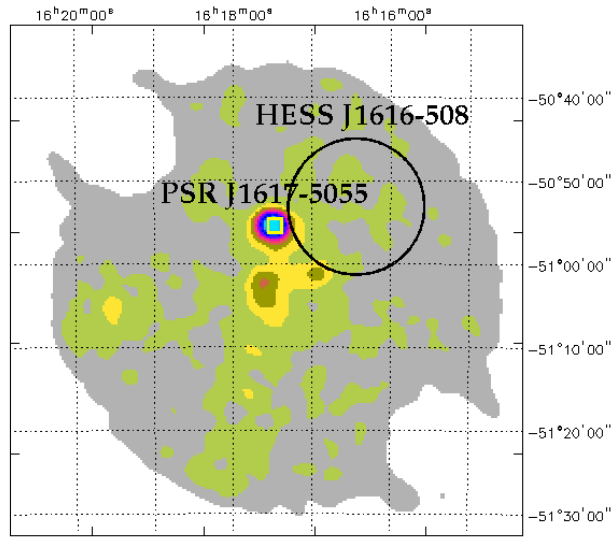}
\caption{MECS 4--10 keV image of the region surrounding 1E 161348--5055.1. The black circle
describes the HESS J1616--508 extension, while the position of PSR J1617--5055 is given
by a box. It is clear that at hard energies the contribution of 1E 161348--5055.1 and
SNR RCW 103 are negligible.}
\label{fig:mecs2}
\end{figure}

\emph{Beppo}SAX pointed at 1E 161348--5055.1 on 17 September 1998, 
23 March and 4 August 1999, for $\sim$16.5, $\sim$38 and $\sim$53 ks, respectively, and
these observations covered, at least in part, the HESS J1616--508 region.
The MECS event files were downloaded from the ASI Scientific Data Centre and analysed
with XIMAGE v4.3.  
Figure~\ref{fig:mecs1} and Figure~\ref{fig:mecs2} show the images of the region 
surrounding RCW 103/1E 161348--5055.1 in the 2--4 and 4--10 keV energy bands;
the \emph{Swift} source was not detected by the MECS instrument probably because it was below its 
detection threshold.

By comparing the MECS images, we found that, while 1E 161340--5055.1 and the 
associated SNR RCW 103 dominate at soft energies, PSR J1617--5055 emerges at high energies; this
evidence confirms the \emph{INTEGRAL} result, and points to the pulsar as the only
source emitting above a few keV. However, this also implies 
that it is difficult to examine the energy spectrum of PSR J1617--5055 in the soft energy band where
the diffuse emission from RCW 103 contributes significantly.
Therefore, in all three \emph{Beppo}SAX observations of PSR J1617--5055,
events for spectral analysis were selected in circular regions of 
3$^{\prime}$ in the 3.5--10 keV energy band. Background spectra were extracted in a region
far from the SNR RCW 103, so that the contamination from its diffuse emission is negligible.
For this off-axis source, we used the appropriate MECS ancillary response
files to correct for the effects of vignetting.

We find that an absorbed power law, in which the intrinsic column density is left free to vary, 
provides a good description of the MECS data for all three observations. 
For each individual measurement, spectral parameters are
compatible within their uncertainties and no flux variation is observed between 
measurements. In view of these indications, we summed together the three datasets 
and estimated the average spectral properties. 
The absorbed power law remains the best-fit model for the pulsar emission
($\chi^{2}/\nu=24.2/24$), giving 
$N_{\rm H} = \left(2.9^{+5.3}_{-2.9}\right)\times10^{22}$ cm$^{-2}$, $\Gamma=1.63^{+0.42}_{-0.34}$ and an 
average 
3.5--10 keV observed flux of $\sim$$2.6\times 10^{-12}$ erg cm$^{-2}$ s$^{-1}$. 
Our spectral parameters are compatible, within their respective uncertainties, with the \emph{ASCA} 
results (Torii et al. 1998, 2000).

The extrapolation of the 3.5--10 keV spectrum of PSR J1617--5055 at high energies 
is also in agreement, within uncertainties, with the \emph{INTEGRAL} flux.

Overall, the \emph{Beppo}SAX data indicate that above a few keV (i.e. in the non-thermal regime)
only one source is bright and stable, PSR J1617--5055.
Its spectrum is hard and absorbed as expected from a pulsar located in the galactic plane;
as the Galactic column density in the direction of the source is $2.4\times10^{22}$ cm$^{-2}$ (Dickey \& 
Lockman 1990), we conclude that the measured $N_{\rm H}$ is not intrinsic, but likely
due to Galactic interstellar absorption.

\section{\emph{XMM-Newton} observation and data analysis}

\begin{figure}
\includegraphics[width=1.0\linewidth]{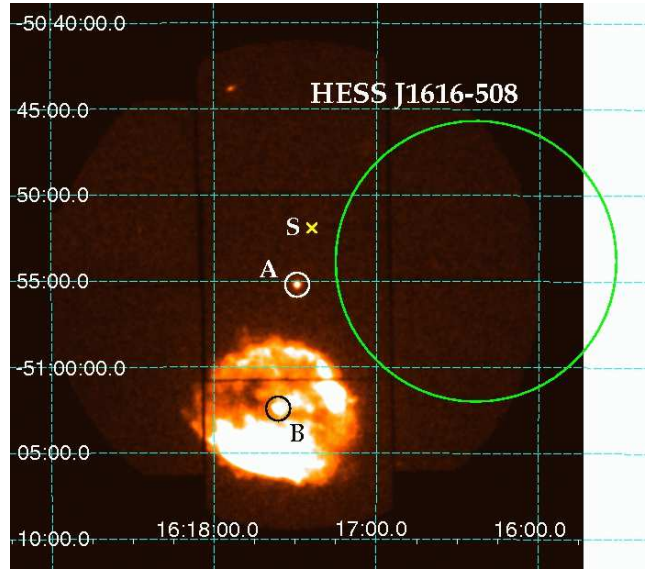}
\caption{\emph{XMM-Newton} MOS 0.5--10 keV image of the region surrounding 1E 161348--5055.1. 
The green circle describes the HESS J1616--508 extension.
PSR J1617--5055 and 1E 161348--5055.1 (circles labelled as A and B respectively) 
are the only sources detected, using {\sc emldetect} task, at energies above 4 keV (see text).
We also plot the position (boxes labelled as 1 and 2)
of the two sources detected within the XRT field of view.
The source labelled as S (yellow cross) is associated to the star HD 146184.}
\label{fig:xmm}
\end{figure}

\begin{figure}
\includegraphics[width=0.75\linewidth,angle=-90]{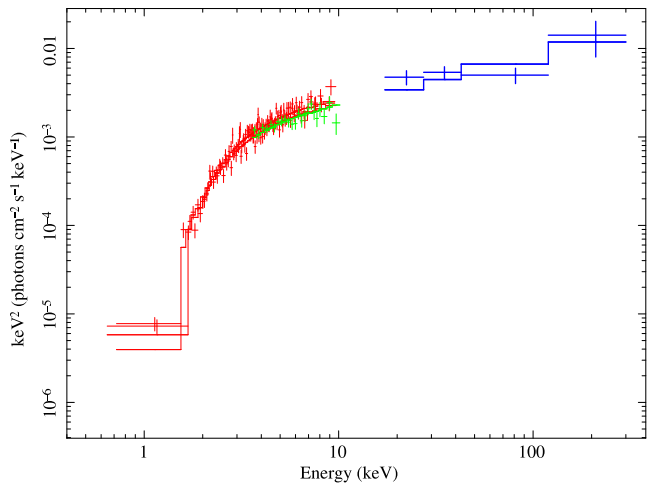}
\caption{Broad band spectrum of PSR J1617--5055 fitted with an absorbed power law. Data are from
\emph{XMM-Newton}/MOS (red), \emph{Beppo}SAX/MECS (green) and \emph{INTEGRAL} IBIS/ISGRI (blue).}
\label{fig:broadspec}
\end{figure}

To complete the picture of the region of interest here, we have also reanalysed  \emph{XMM-Newton} 
data, which have already been discussed in various papers (Matsumoto et al. 2007; De Luca et al. 2006;
Becker \& Aschenback 2002) and which were collected during a 28 ks observation  
made on 2001 September 3 (obs ID: 0113050701). First, we searched for X-ray point sources in the MOS images. 
Apart from RCW 103, PSR J1617--5055 and 1E 161348--5055.1, we found a number of other objects,
none of which fall inside the HESS extension (see Figure~\ref{fig:xmm}); this is not in contradiction 
with the \emph{Swift}/XRT result as the only source seen by this instrument is outside the 
\emph{XMM-Newton} field of view.
The closest new X-ray source to HESS J1616--508 is located at R.A.=16$^{\rm h}$17$^{\rm m}$ 23$^{\rm 
s}$.87 and Dec=--50$^{\circ}$ 51$^{\prime}$ 52$^{\prime \prime}$.48
(J2000.0) with a positional uncertainty of 1$^{\prime \prime}$ 
and is associated to a star (HD 146184).
When we looked at the MOS images, obtained in the 4--10 keV energy band,
only two sources are clearly detected, PSR J1617--5055 and 1E 161348--5055.1.
Contrary to what was found for \emph{Beppo}SAX, the \emph{XMM-Newton} flux for 
1E 161348--5055.1 is now comparable with that of the pulsar, clearly indicating
that 1E 161348--5055.1 is highly variable.
Indeed, De Luca et al. (2006) found variability both in flux and spectral
shape: in particular the spectrum is steep ($\Gamma \sim3$) when the source is bright and 
it is hard when dim.
All these characteristics (compactness, variability and spectral shape) suggest that 1E 
161348--5055.1
is unlikely to be associated with HESS J1616--508, which is not variable and is extended, leaving as 
an alternative possible counterpart only the pulsar.
We have therefore reanalysed the MOS spectral data of this object with the aim of combining 
them with \emph{Beppo}SAX and \emph{INTEGRAL} observations.
Source spectra have been extracted from a circular region of
$30^{\prime \prime}$ radius, while background spectra were taken
from source-free circular regions of $20^{\prime \prime}$ radius around the source and in the same CCD.
Spectral fits were performed simultaneously with MOS1 and MOS2 in the
0.5--10 keV energy band. The cross-calibration constant MOS2/MOS1 was left free to vary and found 
to lie in the range 1.06--1.16.
We find that the \emph{XMM-Newton} spectrum is well described by an absorbed power law having 
parameters compatible with those reported by Becker \& Aschenbach (2002).
When these data are combined with the \emph{Beppo}SAX/MECS and \emph{INTEGRAL} data, we obtained a similar 
spectral shape with the following parameters: 
$\Gamma=1.42_{-0.10}^{+0.12}$, a column density $N_{\rm H}=\left(3.87_{-0.28}^{+0.36} \right) 
\times10^{22}$
cm$^{-2}$, an unabsorbed 2--10 keV flux of $4.2\times 10^{-12}$ erg s$^{-1}$ cm$^{-2}$,
and cross calibration constants MECS/MOS and IBIS/MOS in the range 0.91--1.03 and 
0.65--1.22, respectively.  
These combined data are shown in Figure~\ref{fig:broadspec}: 
the good match between the various datasets and also the hard shape of the pulsar 
spectrum are evident from the figure. It is worth noting that this broad band spectrum is 
the result of two components, pulsed and unpulsed, with the latter emission being possibly associated to an 
as yet unseen PWN. 
Indeed, Becker \& Aschenbach (2002) found that the pulsed fraction contributes to $\sim$50$\%$ in the 
2.5--15 keV band,
with some indication (albeit not statistically significant) of an increase of the pulsed fraction with energy.

To look for this PWN, we have also searched the MOS 0.5--4 keV image  
for the presence of diffuse emission around the pulsar, taking 
into account an annular region having internal and external radius of 40$^{\prime \prime}$
and 120$^{\prime \prime}$ and a total area of 11.2 arcmin$^{2}$; a similar region was used to estimate 
the background. We found a count rate of $(394\pm6)$
and $(243\pm5)$ counts arcmin$^{-2}$ for the source and the background, 
respectively. This implies that the contribution of the diffuse emission is 
$(151\pm8)$ counts arcmin$^{-2}$, which corresponds to a 19$\sigma$ detection.
In the 4--10 keV energy band the count rates become 
$(130\pm3)$ and $(117\pm3)$ counts arcmin$^{-2}$ again for the source and the
background, respectively; thus the contribution of this diffuse component becomes less
evident above 4 keV ($13\pm4$ counts arcmin$^{-2}$) but still detectable, albeit at
only 3$\sigma$ level.
We have also checked for the presence of any asymmetry in this diffuse emission
(i.e. looking in the direction towards and away from the HESS source), but we were 
unable to find any significant evidence for it.
It is difficult, with the present data, to assess if this diffuse component is 
due to the presence of a PWN around PSR J1617--5055 or to emission related to RCW 103, especially in view 
of the \emph{Suzaku} detection of soft thermal emission near the location of HESS J1616--508 
(Matsumoto  et al. 2007).
Taking the \emph{XMM-Newton} data at face value and fitting this diffuse emission with an absorbed power 
law ($\Gamma = 2$ and $N_{\rm H}= 3 \times10^{22}$ cm$^{-2}$),
we estimate a 2--10 keV observed flux of $1.3\times 10^{-12}$ erg cm$^{-2}$ s$^{-1}$,
which can be used as an upper limit to the presence of a PWN around PSR J1617--5055.  
The value we measure is much higher than the thermal emission flux seen by \emph{Suzaku} and it is also 
above the upper limit to the hard X-ray emission from HESS J1616--508 set by \emph{XMM-Newton}.
This, coupled with the possible detection of hard emission above 4 keV, suggests the 
possibility that we are indeed
observing the PWN associated with PSR J1617--5055; it follows that its dimension needs 
to be similar to that of the anular region used to estimate the diffuse component ($\sim$  
120$^{\prime \prime}$). 
Note, however, that radio data show no evidence, in the pulsar direction, for extended emission 
on spatial scale $>$ 6$^{\prime \prime}$ down to a 3$\sigma$ limiting flux of 1.2 mJy (Kaspi et al. 1998). 
An alternative scenario is that the PWN is closer to the pulsar and its emission is 
mixed with that of PSR J1617--5055; in this case the PWN 2--10 keV flux is $\le$ 2$\times 10^{-12}$ 
erg cm$^{-2}$ s$^{-1}$ ( i.e. 50$\%$ of the total 
flux measured by \emph{XMM-Newton}/MOS and \emph{Beppo}SAX/MECS) and its dimension 
$\le$ 30$^{\prime \prime}$ (considering the extraction radius used for the \emph{XMM-Newton}/MOS). 
We consider the above flux as an upper limit 
because diffuse emission, unrelated to the PWN, could also contribute here.

\section{Discussion}

A detailed X-ray analysis of the region surrounding HESS J1616--508 provides 
one clear result: there is no totally convincing X-ray counterpart found 
within the TeV extension.
Searches in various radio archives~(http://heasarc.gsfc.nasa.gov/docs/archive.html) 
give a similar result, thus 
suggesting that either a ``dark particle accelerator" (an object with no X-ray/radio 
counterpart) is producing very high energy $\gamma$-rays or that any plausible 
counterpart is very weak (if positionally coincident with the HESS object) or  
located outside the TeV extension.
Indeed, the first possibility was presented and discussed by the \emph{Suzaku} team 
(Matsumoto et al. 2007) on the basis of their results. 
Here, we would like to suggest the alternative scenario derived from our data set.

\emph{Swift}/XRT finds only one X-ray source, but the low statistical quality of the data
and its characteristics make it an unlikely candidate for the source that powers
the HESS nebula.
Therefore, the only other possibility is to explore energetic objects located 
in the region surrounding the HESS source. This region is a zone of complex stellar activity 
due to its location 
tangential to our line of sight to the Norma spiral arm. There are two nearby SNRs 
(RCW 103 and Kes 32) and three pulsars (PSR J1617--5055, PSR J1616--5109 and PSR J1614--5048)
(Vink 2004; Torii et al. 1998). Of the two SNR, only RCW 103 is of some interest mainly due to the 
presence of the compact source 1E 161348--5055.1 at its centre: however, this object is 
too variable in X-rays to be possibly associated with the stable HESS source.
The relevant characteristics of the three pulsars in the region are presented in Table~\ref{Tab1}. 
Here the final column shows 
the TeV luminosity as a percentage of the pulsar spin down energy loss if situated 
at the distance of each pulsar.
Whereas in principle, the total energy available for the TeV emission could possibly extend up to 
a maximum of $\dot{E}$ integrated over the lifetimes of the electrons against Compton losses
($\sim$5 kyr), observations of pulsar-driven HESS sources (Aharonian et al. 2006a) have shown that the 
TeV emission is typically a few percent of the instantaneous energy loss of the pulsar. This fact, coupled 
with the \emph{INTEGRAL} detection of only PSR J1617--5055 in the 
region, makes this young 69 ms pulsar the only realistic source of energetic $\gamma$-rays.

\begin{table*}
\begin{center}
\caption{Characteristics of the pulsar located nearby the HESS extension}
\label{Tab1}
\begin{tabular}{lcccccc}
\hline
\hline
 Pulsar & Period & $\dot{P}$ & Energy loss&  Characteristic age &
Distance & $L_{\rm HESS}$ \\
   &(ms)  & (s s$^{-1}$ $\times10^{-13}$) &(erg s$^{-1}$)& (yr) & (kpc)& ($\%$ $\dot{E}$) \\
\hline
\hline
PSR J1617--5055 &  69  & 1.35 & $1.6\times10^{37}$ & 8130 & 6.46 & 1.35 \\
PSR J1616--5109 & 1220 & 0.19 & $4.2\times10^{32}$ & $1.01\times10^{6}$ & 18.97 & $>$ 100   \\
PSR J1614--5048 & 231  & 4.95 & $1.6\times10^{36}$ & 7400 & 7.24 & 17 \\
\hline
\end{tabular}
\end{center}
\end{table*}

\subsection{PSR J1617--5055 as a likely counterpart}

PSR J1617--5055 has been well studied both at X-ray and radio frequencies (Torii et al. 1998, 2000;
Kaspi et al. 1998) and its parameters estimated with some accuracy. From the measured $\dot{P}$, a
surface magnetic field of $3.1\times10^{12}$ Gauss and a spin-down luminosity $1.6\times10^{37}$ 
erg s$^{-1}$ have
been evaluated; the implied characteristic age for PSR J1617--5055 is 8.1 kyr, which makes PSR
J1617--5055 one of the youngest of all known pulsars (Kaspi et al. 1998). 

The radio dispersion estimate provides a distance in the range 6.1--6.9 kpc, compatible with the
column density measured in the source direction. The most likely companion SNR to PSR J1617--5055 is RCW
103, however a clear mismatch both in terms of distances and ages argues against an
association between the two (see for example Kaspi et al. 1998); furthermore, the required pulsar
velocity in case of an
association would have to be very high, in clear disagreement with the lack of a visible forward
bow shock nebula (Gaensler et al. 2004) or of a perturbation in the
structure of the SNR.
Clearly, an open question remains on where the SNR left behind by the
birth of this young pulsar is.
As there is no obvious SNR association, the only possible site where the TeV
photons may be generated is the pulsar itself or an associated PWN/jet.
In the first case, a model like the one proposed by
Hirotani (2001) could be at work in the source,
but it is then difficult to explain the offset observed between the
X/$\gamma$-ray emission and the TeV one.
Alternatively, the energetic source of high energy electrons could be a
jet from the pulsar or an as-yet unseen PWN.

Using \emph{XMM-Newton} observational data we have performed a careful
analysis of the X-ray image to search for any spatial extension/asymmetry of the emission
around the pulsar. Some
evidence for extended X-ray emission was found over the 0.5--10 keV range although this could not
be firmly associated
to a PWN due to the presence of the nearby SNR RCW 103. Despite
this uncertainty, some information could
be gained from the data: in fact, the X/$\gamma$-ray and TeV emission are
different in size (by a factor of less than 4--16, depending on which of the two PWN scenarios 
described in Section 5 we adopt)
and offset from each other by $\sim$10$^{\prime}$; if the PWN is the production
site of the TeV photons, then we also have a difference in flux greater than a factor of 20--30.
A natural way to explain the different sizes of the emitting regions for the TeV photons as
compared to the X/soft $\gamma$-ray ones is related to the different cooling
time-scales and hence
propagation distances: electrons producing photons of different energies
spend different periods to emit them over different emitting volumes.
The lifetimes of the electrons that generate the TeV photons against
Compton cooling in the Cosmic Microwave Background (CMB) is $\sim$5000 years and they are able to 
propagate further into the interstellar medium; even if there is a significant magnetic field
present in the region making their lifetime shorter, it is still
$\sim$3 kyr for a 10 $\mu$Gauss field. The lifetimes of the more energetic
electrons that produce
the X/$\gamma$-ray fluxes by the synchrotron process in the same magnetic
field is of the order of a few hundred years.
If we simplistically assume that the relative size of the X and
$\gamma$-ray objects are
proportional to the lifetimes of the parent electrons then we would
expect, for a 16$^{\prime}$
TeV source, to see a much smaller sized object ($\sim$1$^{\prime}$) in 
the X/$\gamma$-ray domain.
For continuous injection and radiative losses that are short
compared to the age of the source, the spectral index of the electrons
steepens with a
corresponding steepening of the photon emission. The observed spectral
index of the combined
X/soft $\gamma$-ray emission ($\Gamma=1.4$), and that of the TeV emission
($\Gamma=2.4$) are
consistent with this scenario. The X/soft $\gamma$-ray fluxes represent
the uncooled component
and the TeV photons are derived from inverse Compton scatters of the synchrotron
cooled electrons on the
CMB: we may also expect the cooled electrons, which produce the TeV photons,
to create an extended EUV/soft X-ray fluxes through the synchrotron process.
However, in view of the large column density observed in the source
direction, it is unlikely that we would be able to detect such a weak
diffuse UV source. In fact,
we have inspected the available \emph{Swift}/UVOT images, but we failed to
detect this emission, due to the combination of the strong
absorption along the line of sight,
and contamination by a significant number of bright point sources along the
same line of sight. While overall the scenario we described is compatible
with the observational evidence, it is not clear why the electrons should stream
out on one side only to generate their high energy inverse Compton
$\gamma$-ray photons. However, despite a lack of understanding of this phenomenon,
an increasing number
of TeV sources are found to be shifted away from the pulsars that
generated them.

Perhaps the most prominent example of this new class of TeV objects is
HESS J1825--137 (Aharonian et al. 2005b).
This source has many similar
characteristics to HESS J1616--508: an associated pulsar of comparative
period (101 ms) and age ($2\times10^{4}$ yr) that is coincident with an X-ray and possibly
with a soft $\gamma$-ray source (Malizia 2006, private communication);
the TeV luminosity is a similar fraction (1--2\%) of the pulsar spin down
power; the TeV emissions have similar photon indices of $\sim$2.4;
both TeV nebulae are similar in size (within a factor of 2) and offset
from the position of the pulsar by the same amount; in both objects the
TeV emitting region is larger that the X-ray one; finally, no SNR has been detected in either
case. However, whereas HESS J1825--137 does have an offset PWN (G18.0--0.7,
Gaensler et al. 2003) located to
the south, the presence of a visible PWN in HESS J1616--508 is highly uncertain.
Taking advantage of this similarity,
we have estimated the X-ray flux of the eventual PWN around PSR
J1617--5055, using the
information available for PWN G18.0--0.7, i.e. the same spectrum
(see Gaensler et al. 2003), but scaled by an appropriate
factor to take into account the differences in distance and column
density. We estimated a 2--10 keV observed flux of $\sim$2 $\times10^{-13}$ erg cm$^{-2}$
s$^{-1}$, far below
the flux value estimated for the two PWN scenarios. If the PWN is more extended, 
this flux mismatch points to the possibility that at least part of the diffuse emission we 
measured is due to a PWN; in the alternative case of a PWN which is more compact and mixed with the pulsar 
emission, the flux difference implies than only a fraction of the unpulsed component is due to 
the plerion emission. This leaves both scenarios as viable, although still rather undefined. 
Clearly, the location of PSR J1617--5055 near RCW 103, the high column density in the source 
direction,
as well as the geometry/intensity of any PWN present in the source, make the detection of such 
a component very difficult to achieve. Clearly, a long dedicated X-ray observation to disintangle 
these various components is required to settle this issue.
Concerning the offset morphology, asymmetric reverse shock interactions have
been proposed for the case of the Vela system (HESS J0835--455) on the
basis of hydro-dynamical simulations (Blondin et al. 2001) and
subsequently observed in \emph{Chandra} images (Helfand et al. 2001).
The X-ray images of the diffuse emission associated with
HESS J1825--137 (Gaensler et al. 2003) are also asymmetric with respect to
the position of the pulsar but there is no evidence of a one-sided
reverse shock process taking place. Although in PSR J1617--5055  there is
no evidence of the existence of a PWN, let alone of any one-sided reverse shock, this does
not necessarily dissociate it from HESS J1616--508. While much more understanding of the
processes at work in TeV sources is needed to explain all the different morphologies
observed, we cannot ignore the fact that HESS J1616--508 is close to a pulsar 
energetic enough to explain the TeV flux by its spin down power.

\section{Summary and Conclusions}

Two conclusions may be drawn directly from the observations presented in this work.
First, that the power source that fuels the X/$\gamma$-ray emission detected
by \emph{INTEGRAL}, \emph{XMM-Newton} and \emph{Beppo}SAX
is almost certainly derived from the pulsar PSR J1617--5055. Apart from the
positional justification, this concept is fully consistent with the timing evidence coming from the
\emph{XMM-Newton} data (Becker \& Aschenbach 2002) and the expected energetics of the system
(see Table~\ref{Tab1}).
The observed 2--10 and 20--100 keV fluxes of $4.2 \times10^{-12}$ erg cm$^{-2}$ s$^{-1}$
and $1.37 \times10^{-11}$ erg cm$^{-2}$ s$^{-1}$ translate, at the distance of
the pulsar, to luminosities which are approximately $\sim$$2\times10^{34}$ erg s$^{-1}$
and $\sim$$7\times10^{34}$ erg s$^{-1}$, or 0.1\% and 0.4$\%$ of
the spin down losses. These values are fully within
the range observed from young radio pulsars like the Crab and from PWN systems observed by
\emph{INTEGRAL} (Dean et al. in preparation). Likewise, for the pulsar to be responsible for the flux at
TeV energies, 1.2$\%$ of the spin down energy loss is needed, in agreement with that
observed in other HESS sources.
The fact that \emph{INTEGRAL} finds a point-like source at the position of PSR J1617--5055, that its 
spectrum smoothly connects to \emph{Beppo}SAX and \emph{XMM-Newton} spectra, and that the luminosity has a 
value consistent with what expected from a young radio pulsar, naturally make this pulsar the source that 
powers the high energy emission.
 
Second, that the TeV emission is not positionally coincident with the
pulsar and no other promising candidate counterpart has been found within the TeV extension contour.
However, the lack of precise positional coincidence between HESS J1616--508
and PSR J1617--5055 does not necessarily
dissociate the two objects, particularly in view of recent observational
evidence of TeV sources offset from their powering pulsars. We have argued that 
there is good evidence to assume that we have a
pulsar driven system in which inverse Compton and synchrotron processes combine to
create the observed morphology of a broad-band emitter from keV to TeV energies.
Under the assumption that the HESS photons are a result of electrons
derived from the pulsar
that subsequently suffer Compton scattering with the low energy photons of
the CMB and that
the X- and soft $\gamma$-rays are created through synchrotron losses by
the same (although
more energetic) electron family members in the chaotic magnetic field of
typically 10 $\mu$Gauss
near to the pulsar, then the relative sizes of the VHE and X/$\gamma$
sources are consistent with the expected lifetimes of the parent electrons against 
synchrotron and Compton interactions. The lack of an observable PWN
is intriguing, but not damning, and a rationale for the one-sided nature of
the HESS object not
understood, but not unprecedented (Aharonian et al. 2006b; Mangano et al. 2005).

A long dedicated observation with  \emph{Chandra}
would be highly desirable to settle this issue.

\section*{Acknowledgments}
This research has been supported by ASI under
contracts I/R/046/04 and I/008/07/0. This research has made use of data obtained from SIMBAD
(CDS, Strasbourg, France) and HEASARC (NASA's Goddard Space Flight Centre).


\begin{thebibliography}{99}
\bibitem[\protect\citeauthoryear{Aharonian et al.}{2006a}]{b1} Aharonian, F., Akhperjanian, A. G., 
Bazer-Bachi, A. R. et al., 2006a, ApJ, 636, 777
\bibitem[\protect\citeauthoryear{Aharonian et al.}{2006b}]{b3} Aharonian, F., Akhperjanian, A. G., 
Bazer-Bachi, A. R. et al., 2006b, A\&A, 448, L43
\bibitem[\protect\citeauthoryear{Aharonian et al.}{2005a}]{b4} Aharonian, F., Akhperjanian, A. G., 
Aye, K.-M., et al., 2005a, Science, 307, 1938
\bibitem[\protect\citeauthoryear{Aharonian et al.}{2005b}]{b4bis} Aharonian, F., Akhperjanian, A. G.,
Aye, K.-M., et al., 2005b, A\&A, 442, L25
\bibitem[\protect\citeauthoryear{Becker \& Aschenbach}{2002}]{b5} Becker W., \& Aschenbach B., 2002, 
Proceedings of the 270. WE-Heraeus Seminar on
``Neutron Stars, Pulsars and Supernova Remnants'', eds W. Becker, H. Lesch \& J. Tr{\"u}mper,
MPE Report, Pag. 64
\bibitem[\protect\citeauthoryear{Bird et al.}{2007}]{b6} Bird, A. J., Malizia, A., Bazzano, A., et 
al., 2007, ApJS, 170, 175 
\bibitem[\protect\citeauthoryear{Bird et al.}{2004}]{b7} Bird, A. J., Barlow, E. J., Bassani, L., et al., 
2004, ApJ, 607, L33
\bibitem[\protect\citeauthoryear{Blondin et al.}{2001}]{b8} Blondin, J. M., Chevalier, R. A., \& 
Frierson D. M., 2001, ApJ, 563, 806
\bibitem[\protect\citeauthoryear{De Luca et al.}{2006}]{b12} De Luca, A., Caraveo P. A., 
Mereghetti, S., Tiengo A., \& Bignami, G. F., 2006, Science, 313, 814
\bibitem[\protect\citeauthoryear{Dickey \& Lockman}{1990}]{b13} Dickey, J. M., \& Lockman, F. J., 1990, 
ARA\&A, 28, 215
\bibitem[\protect\citeauthoryear{Gaensler et al.}{2004}]{b15} Gaensler, B. M., van der Swaluw, 
E., Camilo, F., Kaspi, V. M., Baganoff, F. K., Yusef-Zadeh, F., \& Manchester, R. N., 2004, ApJ, 616, 383 
\bibitem[\protect\citeauthoryear{Gaensler et al.}{2003}]{b16} Gaensler, B. M., Schulz, N. S., 
Kaspi, V. M., Pivovaroff, M. J., \& Becker, W. E., 2003, ApJ, 588, 441
\bibitem[\protect\citeauthoryear{Gehrels et al.}{2004}]{b17} Gehrels, N., Chincarini, G., Giommi, 
P. et al., 2004, ApJ, 611, 1005
\bibitem[\protect\citeauthoryear{Goldwurm et al.}{2003}]{b18} Goldwurm, A., David, P., 
Foschini, L., et al., 2003, A\&A, 411, L223
\bibitem[\protect\citeauthoryear{Helfand et al.}{2001}]{b20} Helfand, D. J., Gotthelf, E. V., \& 
Halpern, J. P., 2001, ApJ, 556, 380
\bibitem[\protect\citeauthoryear{Hill et al.}{2004}]{b21} Hill, J. E., Burrows, D. N., Nousek, J. A., 
et al., 2004, Proc. SPIE, 5165, 217
\bibitem[\protect\citeauthoryear{Hirotani}{2001}]{b22} Hirotani, H., 2001, ApJ, 549, 495
\bibitem[\protect\citeauthoryear{Kaspi et al.}{1998}]{b23} Kaspi, V. M., Crawford, F., Manchester, R. N., 
Lyne, A. G., Camilo, F., D'Amico, N., \& Gaensler, B. M., 1998, ApJ, L161
\bibitem[\protect\citeauthoryear{Mangano et al.}{2005}]{b25} Mangano, V., Massaro, E., Bocchino, 
F., Mineo, T., \& Cusumano, G., 2005, A\&A, 436, 917
\bibitem[\protect\citeauthoryear{Matsumoto et al.}{2007}]{b26} Matsumoto, H., Ueno, M., 
Bamba, A., et al., 2007, PASJ, 59, 151
\bibitem[\protect\citeauthoryear{Monet et al.}{1999}]{b28} Monet, A. K. B., Levine, S. E., Monet, D. 
G., Bowell, E. L. G., Koehn, B., \& Bryan, B. 1999, BAAS, 31, 1532
\bibitem[\protect\citeauthoryear{Skrutskie et al.}{2006}]{b30} Skrutskie, M. F., Cutri, R. 
M., Stiening, R., et al., 2006, AJ, 131, 1163
\bibitem[\protect\citeauthoryear{Torii et al.}{2000}]{b31} Torii, K., Gotthelf, E. V., Vasisht, G., 
Dotani, T., \& Kinugasa, K., 2000, ApJ, 534, L71 
\bibitem[\protect\citeauthoryear{Torii et al.}{1998}]{b32} Torii, K., Kinugasa, K., Toneri, T., 
Asanuma, T., Tsunemi H., Dotani, T., Mitsuda, K., Gotthelf, E. V., \& Petre, R., 1998, ApJ, 494, L207
\bibitem[\protect\citeauthoryear{Vink}{2004}]{b33} Vink, J., 2004, ApJ, 604, 693
\bibitem[\protect\citeauthoryear{Winkler}{2003}]{b34} Winkler, C., Courvoisier, T. J.-L., Di Cocco, 
G., 2003, A\&A, 411, L1
\end{thebibliography}
\end{document}